\documentclass[aps,superscriptaddress,eqsecnum,nofootinbib,preprintnumbers,twocolumn]{revtex4}
\pdfoutput=1

\usepackage{longtable}
\usepackage{bm}
\usepackage{relsize}
\usepackage{amsfonts}
\usepackage{amsmath}
\usepackage{amssymb,epsf}
\usepackage{latexsym}
\usepackage{graphicx,epsfig}
\usepackage{amssymb}
\usepackage{float}
\usepackage{subfigure}
\usepackage{epstopdf}
\usepackage[colorlinks=true,citecolor=blue,linkcolor=blue,urlcolor=black]{hyperref}
\usepackage{dcolumn}
\usepackage{psfrag}
\usepackage{wrapfig}
\usepackage{makeidx}
\usepackage{epsf}
\usepackage{color}
\usepackage{multirow}
\usepackage{mathtools}

\begin{document}

\title{Astrometric and polarimetric imprints of hot-spots orbiting parametrized black holes}

\author{João Luís Rosa}
\email{joaoluis92@gmail.com}
\affiliation{Departamento de F\'isica Te\'orica, Universidad Complutense de Madrid, E-28040 Madrid, Spain}
\affiliation{Institute of Physics, University of Tartu, W. Ostwaldi 1, 50411 Tartu, Estonia}

\author{Nicolas Aimar}
\email{ndaimar@fe.up.pt}
\affiliation{Faculdade de Engenharia, Universidade do Porto, s/n, R. Dr. Roberto Frias, 4200-465 Porto, Portugal}
\affiliation{CENTRA, Departamento de F\'isica, Instituto Superior T\'ecnico-IST, Universidade de Lisboa-UL, Avenida Rovisco Pais 1, 1049-001 Lisboa, Portugal}

\author{Diego Rubiera-Garcia} \email{drubiera@ucm.es}
\affiliation{Departamento de F\'isica Te\'orica and IPARCOS,
	Universidad Complutense de Madrid, E-28040 Madrid, Spain}

\date{\today}

\begin{abstract} 
We analyze the observational features of hot-spots orbiting parametrized black hole (BH) spacetimes. We select a total of four BH spacetimes, two from the Johanssen-Psaltis (JP) parametrization, and two from the Konoplya-Zhidenko (KZ) parametrization, corresponding to the most extreme configurations whose shadow sizes are within the $2\sigma$-constraints of the Event Horizon Telescope (EHT). We use the ray-tracing software GYOTO to simulate the orbit of a spherically symmetric hot-spot emitting synchrotron radiation close to a central parametrized BH object, in a vertical magnetic field configuration, and we extract the corresponding astrometric and polarimetric observables for the Stokes parameters I, Q and U, namely the time integrated fluxes, temporal fluxes and magnitudes, temporal centroid, temporal QU-loops, and temporal Electric Field Position Angle (EVPA). Our results indicate that at low inclination the astrometric observables extracted from the parametrized BH spacetimes considered are qualitatively similar to those extracted from the Schwarzschild one, with minor quantitative deviations caused by differences in the size and position of the secondary images. On the other hand, the polarimetric observables at high inclination present qualitative differences, but these are only visible for a short portion of the whole hot-spot orbit. Furthermore, the observables extracted from the JP parametrized BH models deviate more prominently from those of the Schwarzschild model than the ones extracted from the KZ parametrized BH models, with the JP model with a positive free parameter deviating the most among all models tested. Given the strong similarity among the observables extracted from all models tested, we point out that more precise observations are needed to successfully impose constraints on parametrized BH models via this method.
\end{abstract}

\maketitle

\section{Introduction}\label{sec:intro}

Recent advances in gravitational physics have yielded strong observational evidence supporting the existence of ultra-compact objects in the universe. The most significant findings include the detection of gravitational wave signals from the merging of compact-object binaries by the LIGO-Virgo-KAGRA (LVK) collaboration \cite{LIGOScientific:2016aoc,KAGRA:2021vkt}, the imaging of shadow-like regions close to the centers of the M87 \cite{EventHorizonTelescope:2019dse} and the Milky Way (at Sgr A* \cite{EventHorizonTelescope:2022wkp}) galaxies, by the Event Horizon Telescope (EHT) collaboration, and the detection of orbiting infrared flares close to the galactic center by the GRAVITY collaboration \cite{GRAVITY:2020lpa,GRAVITY2}. These observations align closely with the predictions from Black Hole (BH) spacetimes, namely the Kerr hypothesis \cite{Will:2014kxa,Yagi:2016jml}, which claims that the end-result of full gravitational collapse in any realistic astrophysical scenario is a rotating and (mostly) electrically neutral BH \cite{Kerr:1963ud,Penrose:1964wq}.

In order to test the validity of the Kerr hypothesis, one can verify if the current observations are instead compatible with alternative spacetime geometries. One method to obtain such geometries is to consider additional fundamental fields in General Relativity (GR), from which models can be developed for hairy BHs \cite{Herdeiro:2015waa}, wormholes \cite{Visser:1995cc}, or bosonic stars \cite{Liebling:2012fv}, among several others \cite{Cardoso:2019rvt}. The observational properties of these so-called Exotic Compact Objects (ECOs) have been extensively analyzed in the literature \cite{Eichhorn:2021iwq,Konoplya:2019sns,Eichhorn:2022oma,Rosa:2022tfv,Rosa:2024eva,Rosa:2023hfm,Santos:2024vby,Huang:2025xqd,Faraji:2025efg,Rosa:2023qcv,Rosa:2022toh,Rosa:2024bqv,Macedo:2024qky,Aimar:2025uia,Rosa:2025dzq,Tamm:2023wvn}. A different though related method is to consider alternative spacetime geometries that can be developed through modifications of the gravitational theory itself \cite{DeFelice:2010aj,Olmo:2011uz,Clifton:2011jh,Nojiri:2017ncd,Shankaranarayanan:2022wbx,Sotiriou:2008rp}, and the observational properties of these modified BH scenarios have also been widely scrutinized \cite{Uniyal:2022vdu,Hu:2022lek,Pantig:2022gih,Okyay:2021nnh,Yang:2022btw,Heydari-Fard:2023ent,Murk:2024nod,Ovgun:2024zmt,Asukula:2023akj}. Nevertheless, the development of BHs and ECOs models in modified theories of gravity remains a typically difficult task, due to the more complicated field equations arising in these theories.

An alternative to obtaining spacetime geometries through directly solving the field equations in modified theories of gravity is the employment of parametrized solutions, which allows one to remain theory-agnostic while studying families of spacetime geometries containing an arbitrary number of free parameters. Several parametrization methods have been proposed, most notably the Johanssen-Psaltis (JP) \cite{Johannsen:2011dh}, the Konoplya-Zhidenko (KZ) \cite{Konoplya:2016pmh} (a restriction of the parametrization  proposed in \cite{Konoplya:2016jvv}), and the Rezzola-Zidenko (RZ) \cite{Rezzolla:2014mua} parametrizations. Scattering and grey-body factors \cite{Magalhaes:2022pgl,Dubinsky:2024nzo,Konoplya:2024vuj}, no-hair theorem \cite{Khodadi:2021gbc}, quasi-normal modes \cite{Cardoso:2019mqo,Konoplya:2022pbc,Hirano:2024fgp}, shadows \cite{Bambhaniya:2021ybs,Younsi:2021dxe,He:2025rjq,Olmo:2025ctf}, interferometry of higher-order images \cite{Feleppa:2025ejh} and strong-field gravity tests  \cite{Cardenas-Avendano:2019zxd} of several such parametrized solutions have been studied in the literature due to potential differences in the corresponding observables that could act as smoking guns  of the existence of alternative BH configurations to the Kerr solution.

The main aim of this work is to study the astrometric and polarimetric properties of hot-spots orbiting 
two such parametrized (the JP and KZ) BH solutions, as a follow-up of the analysis carried out in \cite{Olmo:2025ctf} on the properties of the photon rings from a thin-accretion disk in time-averaged images.
The study of hot-spots orbiting parametrized BH is motivated by the detection of orbital motion from the flares of Sgr A* \cite{Abuter:2018uum}.
To simulate the observational properties of these parametrized BH models with hot-spots, we make use of the high-precision ray-tracing software GYOTO \cite{Vincent:2011wz,Aimar:2023vcs}, which was proven useful in the imaging of both canonical and alternative BH spacetime metrics in suitable astrophysical scenarios \cite{Vincent:2020dij,Lamy:2018zvj,Vincent:2016sjq}, as well as in the previously mentioned ECO models \cite{Rosa:2023qcv,Rosa:2022toh,Rosa:2024bqv,Macedo:2024qky,Aimar:2025uia,Rosa:2025dzq,Tamm:2023wvn}. This software is adequate for our analysis due to its capability of performing ray-tracing with polarization, which aligns with the most recent measurements of light polarization, i.e., the orientation of the electric field vector, from the EHT \cite{EventHorizonTelescope:2021bee,EventHorizonTelescope:2024hpu,EventHorizonTelescope:2024rju}, GRAVITY \cite{GRAVITY:2023avo}, and ALMA \cite{Wielgus:2022heh} collaborations, from the vicinity of supermassive compact objects. It has been shown that light polarization is sensitive to general relativistic effects \cite{Vincent:2023sbw,Himwich:2020msm,Palumbo:2023auc}, which emphasizes the adequacy of this framework to test the underlying spacetime geometry.

This work is organized as follows. In Sec. \ref{sec:theory} we introduce the theory behind parametrized BH spacetimes, and specifically the two parametrizations that are used throughout the paper, including the constraints on the model parameters via the EHT observations. In Sec. \ref{sec:polar} we introduce the Stokes parameters, the astrometric and polarimetric observables that can be extracted from the numerical simulations. In Sec. \ref{sec:results} we show our results and analyze how the different models compare based on the observables considered. Finally, in Sec. \ref{sec:concl}, we trace our conclusions.

\section{Background spacetimes}\label{sec:theory}

Parametrized BH solutions are generic solutions for BH spacetimes that are postulated independently of any specific theory of gravity. The spacetime metric that describes these solutions is characterized by an arbitrarily large number of free parameters and, therefore, can potentially accommodate large arrays of specific metrics. For the purpose of this work, we focus on two specific parametrizations, namely the Johanssen-Psaltis (JP) \cite{Johannsen:2011dh} and Konoplya-Zhidenko (KZ) \cite{Konoplya:2016pmh} parametrizations, which we introduce in what follows.

\subsection{JP parametrization}

The JP parametrization is described by a spherically symmetric line element, which in the usual spherical coordinate system $x^\mu=\{t, r, \theta, \phi\}$ takes the form
\begin{equation}\label{eq:JPmetric}
    ds^2=f_\text{S}\left(r\right)\left[1+h_\text{JP}\left(r\right)\right]dt^2+\frac{1+h_\text{JP}\left(r\right)}{f_\text{S}\left(r\right)}dr^2+r^2 d\Omega^2,
\end{equation}
where $f_\text{S}\left(r\right)=1-2M/r$ is the Schwarzschild metric function, $d\Omega^2=d\theta^2+\sin^2\theta d\phi^2$ is the line element on the two-spheres, and $h_\text{JP}\left(r\right)$ is an infinite series of the radial coordinate $r$ of the form
\begin{equation}\label{eq:JPseries}
    h_\text{JP}\left(r\right)=\sum_{k=0}^{\infty}\epsilon_k\left(\frac{M}{r}\right)^k,
\end{equation}
where $M$ is the ADM mass of the spacetime and $\epsilon_k$ are a set of coefficients that characterize the parametrization. The coefficients $\epsilon_0$ and $\epsilon_1$ must vanish in order to preserve the asymptotic flatness of the spacetime, whereas the coefficient $\epsilon_2$ is experimentally strongly  constrained \cite{Williams:2004qba,Bertotti:2003rm}, and thus we set it to zero for our purposes. Following this analysis, the first non-trivial term in the parametrization is proportional to $\epsilon_3\equiv\epsilon$. In this case, a series expansion of the $g_{tt}$ component of the metric at large distances, i.e., $r\gg 1$, takes the form
\begin{equation}\label{eq:JPgtt}
    -g_{tt}\simeq 1-\frac{2M}{r}+\frac{\epsilon M^3}{r^3}+\mathcal O\left(\frac{1}{r^4}\right).
\end{equation}

In a previous work \cite{Olmo:2025ctf}, it was concluded that the parameter $\epsilon$ is constrained by EHT observations to be within the range
\begin{equation}\label{eq:JPbound}
    \epsilon\in\left[-5, 11.4\right],
\end{equation}
which corresponds to the $2\sigma$-constrains on the (indirect) measurement of the size of the shadow's radius via a correlation with the size of the bright ring of radiation that surrounds it \cite{EventHorizonTelescope:2022xqj}, with the lower (upper) bound corresponding to the minimum (maximum) allowed shadow radius. For the purpose of this work, we consider two JP parametrizations for which the value of $\epsilon$ takes the two extreme values of the bound above.

\subsection{KZ parametrization}

The KZ parametrization is also described by a spherically symmetric line element, which in the usual spherical coordinate system takes the form
\begin{equation}\label{eq:KZmetric}
    ds^2=-f_{\text{KZ}}\left(r\right)dt^2+\frac{dr^2}{f_{\text{KZ}}\left(r\right)}+r^2d\Omega^2,
\end{equation}
where the metric function $f_{\text{KZ}}\left(r\right)$ is defined as
\begin{equation}\label{eq:KZfunction}
    f_{\text{KZ}}\left(r\right)=1-\left[1+h_\text{KZ}\left(r\right)\right]\frac{2M}{r},
\end{equation}
and where $h_\text{KZ}\left(r\right)$ is an infinite series of the radial coordinate $r$ of the form
\begin{equation}\label{eq:KZseries}
    h_\text{KZ}\left(r\right)=\frac{1}{2}\sum_{k=0}^{\infty}\eta_k\left(\frac{M}{r}\right)^k,
\end{equation}
where $\eta_k$ are a set of coefficients that characterize the parametrization. Similarly to what happens for the parameters $\epsilon_k$ in the JP parametrization, the parameter $\eta_0$ must vanish in order to preserve the asymptotic flatness of the spacetime, whereas the parameter $\eta_1$ must vanish in order to satisfy the experimental constraints. As such, the first non-trivial term in this parametrization is proportional to $\eta_2\equiv\eta$, and a series expansion of the $g_{tt}$ component of the metric at large distances $r\gg 1$ takes the form
\begin{equation}\label{eq:KZgtt}
    -g_{tt}=1-\frac{2M}{r}-\frac{\eta M^3}{r^3}+\mathcal O\left(\frac{1}{r^4}\right).
\end{equation}
A property of the KZ parametrized spacetimes described by the metric above is that the event horizon is absent when $\eta<-32/27$, and thus these spacetimes correspond to naked singularities instead of BHs. The BH parameter $\eta$ is thus constrained by EHT observations to be within the range
\begin{equation}\label{eq:KZbound}
    \eta\in\left[-\frac{32}{27}, 2\right],
\end{equation}
the lower bound corresponding to the limit at which the object transitions from a BH to a naked singularity, and the upper bound corresponding to the minimum allowed shadow radius. For the purpose of this work, we consider also two KZ parametrizations for which the value of $\eta$ takes the two extreme values of the bound above.

\section{Astrometry and polarimetry}\label{sec:polar}

\subsection{The physics of hot-spots}

Compact flaring events around supermassive black holes, such as those observed for Sgr A$^*$ \cite{Witzel:2018kzq,Witzel:2020yrp},  appear as hotter and brighter-than-average, transient events or {\it hot-spots} \cite{Broderick:2005jj,Hamaus:2008yw,Yfantis:2024eab}.  It has been shown \cite{Abuter:2018uum,Wielgus:2022heh,GRAVITY:2023avo} that these hot-spots orbit the inner region of the accretion disk around the BH by the extreme bending near their photon spheres, emitting radiation which suffers Dopper-boost, gravitational redshift, and gravity-lensing, potentially revealing information about the BH nature and the disk's orientation. For the sake of this work we simulate the hot-spot as a spherically symmetric, localized object emitting synchrotron radiation while on its circular orbit around the parametrized BH, in a vertical magnetic field configuration as follows from the measurements of \cite{Wielgus:2022heh,GRAVITY:2023avo}. The choice of synchroton radiation is anchored on the grounds of its relevance as an emission mechanism generated by the magnetized, hot gas surrounding the BH and characterized by broadband, highly beamed and polarized electromagnetic radiation \cite{Rybicki}. The setting will provide us with polarimetric images via several observables \cite{Vos:2022yij} (as described next), which is a different probe for near-horizon black hole physics, complementary  to that obtained from time-averaged images of the accretion flow (i.e. the canonical BH shadows), and potentially acting as a way to test new physics in strong-gravity regimes \cite{Shahzadi:2022rzq}.

\subsection{Stokes parameters}

To analyze the astrometric and polarimetric imprints of the parametrized BH spacetimes introduced above, we make use of the ray-tracing software GYOTO \cite{Vincent:2011wz,Aimar:2023vcs}. The software produces a set of two-dimensional matrices representing the specific intensities of the Stokes parameters, namely the total intensity $I$ and the Stokes $Q$ and $U$, which are of interest in this work. These Stokes parameters can be introduced as follows (for further details we refer the reader to \cite{Vincent:2023sbw}). The electric field vector of an incident wave on the observer's screen is (bold format stands for vectors)
\begin{equation}
    \textbf{E}=E\left(\cos\chi_o\textbf{e}_\alpha +\sin\chi_o\textbf{e}_\beta\right),
\end{equation}
where $E$ is the amplitude of the electric field, $\chi_o$ is the observed Electric Vector Position Angle (EVPA), and the vectors $\textbf{e}_\alpha$ and $\textbf{e}_\beta$ form an orthonormal basis in the plane of the observer's screen. Given the fact that the EVPA (which contains the information about the polarization) is not directly observable, we first define the Stokes parameters as
\begin{eqnarray}
    I&=&E^2,\nonumber \\
    Q&=&I\cos\left(2\chi_o\right), \\
    U&=&I\sin\left(2\chi_o\right).\nonumber
\end{eqnarray}
Given the definitions of $Q$ and $U$ above, the observed EVPA can be calculated as
\begin{equation}
    \chi_o=\frac{1}{2}\text{atan2}\left(Q, U\right),
\end{equation}
being thus constrained to the interval $\chi_o\in\left[-\frac{\pi}{2},\frac{\pi}{2}\right]$.

For the purpose of this work, we consider the emission mechanism to be synchrotron radiation. The polarization vector for synchrotron radiation $\textbf{f}_e$ in the rest-frame of the emitter is orthogonal to both the wave vector $\textbf{K}_e$ and the magnetic field vector $\textbf{B}_e$. The relationship between these three vectors is \cite{RL}
\begin{equation}
    \textbf{f}_e=\textbf{K}_e\times\textbf{B}_e.
\end{equation}
The polarization vector is always orthogonal to the direction of light propagation in vacuum, which follows null geodesics. The parallel transport of this vector along null geodesics is done numerically with GYOTO.

\subsection{Observables}

The Stokes parameters output by the GYOTO software are in the form of two-dimensional matrices of specific intensities $S_{lm}$, where $S$ generally represents each of the Stokes parameters $I$, $Q$, and $U$, for each instant of time $t_k$. The indices $l$ and $m$ represent the pixels on the observer's screen. The simulation is repeated for several time instants $t_k\in\left[0, T\right[$, where $T$ is the orbital period of the emitting source. This results in a cube of data $S_{klm}$. The following astrometric and polarimetric observables can be generated from the cube of data: \\

\paragraph{Time integrated flux}
\begin{equation}
    \left<S\right>_{lm}=\sum_k S_{klm},
\end{equation}
\paragraph{Total temporal flux}
\begin{equation}
    F^S_k = \sum_{l}\sum_{m}\Delta\Omega S_{klm},
\end{equation}
\paragraph{Total temporal magnitude}
\begin{equation}
    m^S_k=-2.5\log\left(\frac{F^S_k}{F_\text{min}}\right),
\end{equation}
\paragraph{Temporal centroid}
\begin{equation}
    \vec{c}^S_k=\frac{1}{F^S_k}\sum_l\sum_m\Delta\Omega S_{klm}\vec{r}_{lm},
\end{equation}
\paragraph{Temporal QU-loops}
\begin{equation}
    \vec{L}_k=\frac{1}{F_k^I}\left(F_k^Q\vec{e}_Q+F_k^U\vec{e}_U\right)
\end{equation}
\paragraph{Temporal EVPA}
\begin{equation}
    \chi_k=\frac{1}{2}\text{atan}\left(F_k^Q, F_k^U\right),
\end{equation}
where $\Delta\Omega$ is the solid angle of a single pixel on the observer's screen, and $\vec{r}_{lm}$ is the position vector of the pixel $\{l,m\}$ on the observer's screen, the vectors $\vec{e}_Q$ and $\vec{e}_U$ form an orthonormal basis in the QU-plane, and $F_\text{min}$ represents the minimum value of the flux across all models, to guarantee that all images are normalized to the same value and to provide a ground for comparison. In this work, we present the time integrated flux for the three Stokes parameters $I$, $Q$ and $U$, we present the temporal centroid and the temporal magnitude for the Stokes parameter $I$ only, and we present the total temporal flux for the normalized Stokes parameters $Q/I$ and $U/I$ only.

\subsection{Numerical setup}

In the GYOTO software, we simulate the orbit of a spherically symmetric light source with a radius $R_s=0.5M$ around a central compact object described by each of the four parametrized BH solutions introduced in Sec. \ref{sec:theory}, which we assume as an alternative modeling for Sgr A$^*$. The source is set to orbit in the equatorial plane $\theta=\pi/2$, with a constant orbital radius $r_o=8M$ and with Keplerian velocity. The source emits synchrotron radiation from a fidutial thermal distribution of electrons, with a dimensionless temperature $\Theta_e=200$ and number density $n=6.6 \text{cm}^{-3}$, with a frequency of $f=230\text{GHz}$. The magnetic field configuration is vertical with a field strength of $B\simeq 0.34\text{G}$, corresponding to a magnetization parameter of $\sigma=0.001$, in accordance with experimental measurements \cite{Wielgus:2022heh,GRAVITY:2023avo}. The ADM mass of the central object is $M=4.2\times 10^6 M_\odot$, where $M_\odot$ is the solar mass, and the distance from the central object to the observer's screen is $d=8.25\text{kpc}$. We generate images with a resolution of $1000\times1000$ pixels, with a field of view of the observer of $250 \mu\text{as}$, and for two observation inclinations, namely $20^\circ$ and $80^\circ$.

\section{Results and discussion}\label{sec:results}

\subsection{Overall view}

\begin{figure*}
    \centering
    \includegraphics[width=0.95\linewidth]{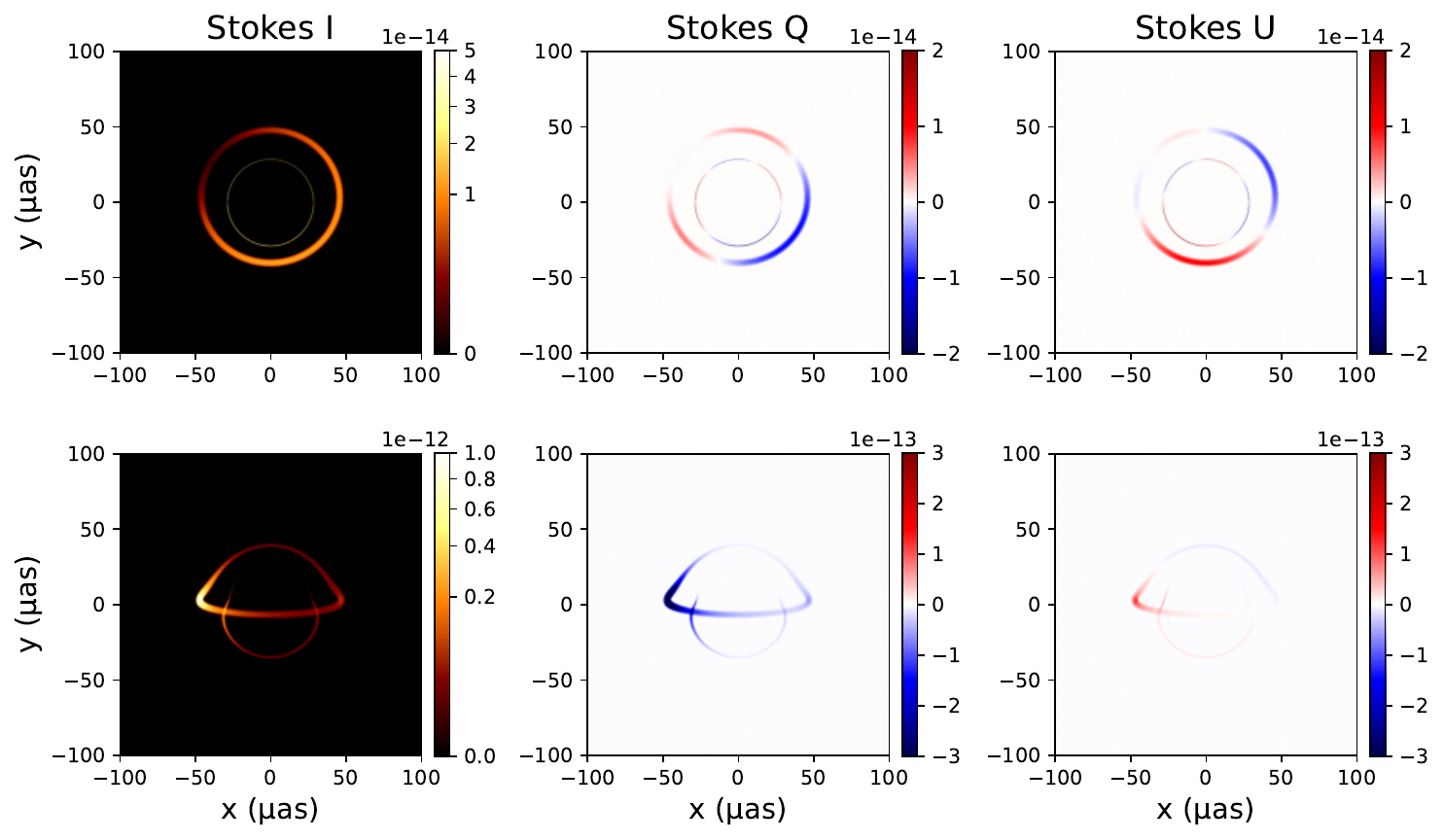}
    \caption{Integrated flux for the Stokes $I$ (left column), Stokes $Q$ (middle column), and Stokes $U$ (right column) parameters, for an observation inclination of $20^\circ$ (top row) and $80^\circ$ (bottom row), for the Schwarzschild BH.}
    \label{fig:integrated_SBH}
\end{figure*}

\begin{figure*}
    \centering
    \includegraphics[width=0.95\linewidth]{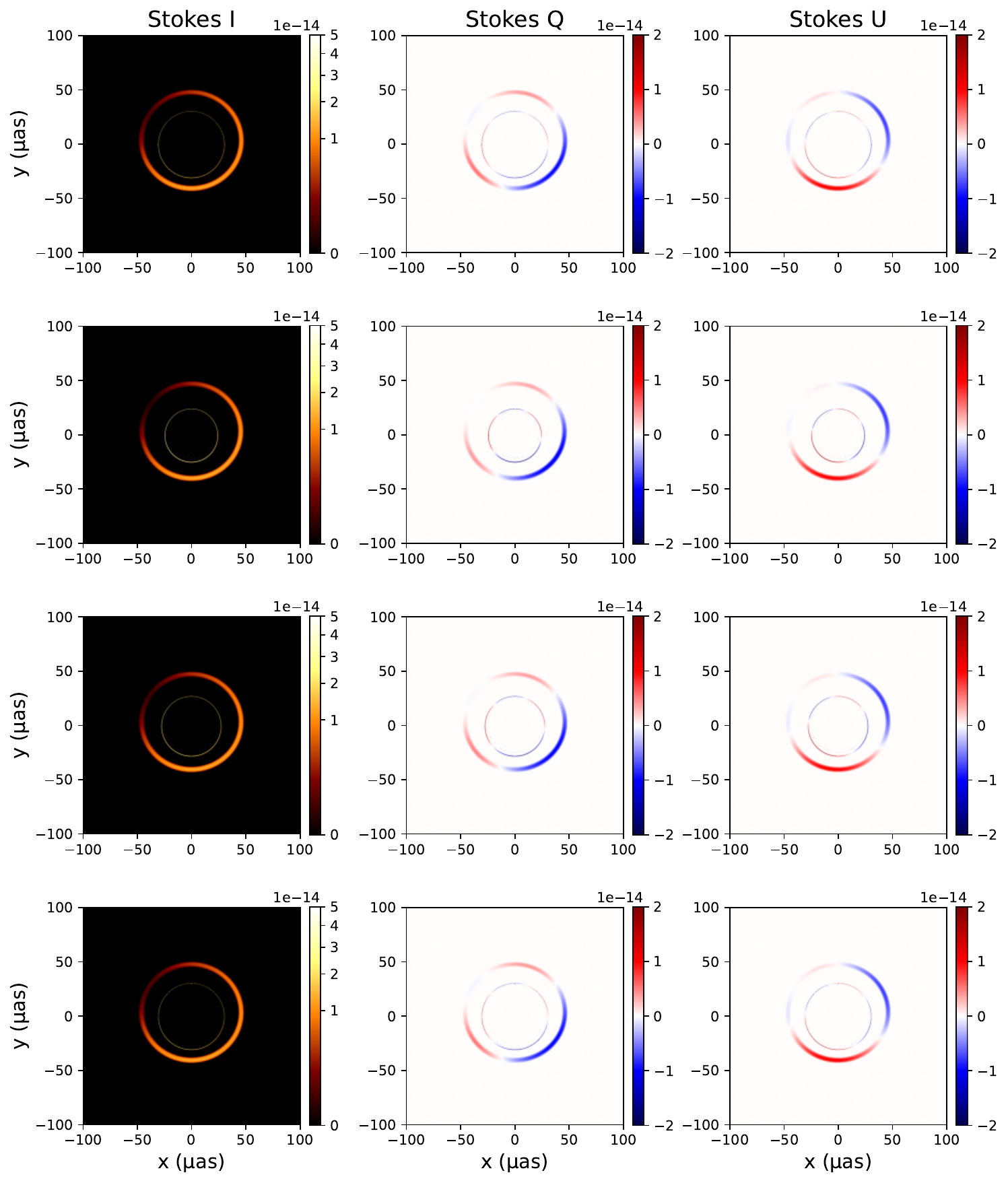}
    \caption{Integrated flux for the Stokes $I$ (left column), Stokes $Q$ (middle column), and Stokes $U$ (right column) parameters, for an observation inclination of $20^\circ$, for the negative JP model (first row), positive JP model (second row), negative KZ model (third row), and positive KZ model (fourth row).}
    \label{fig:integrated_20}
\end{figure*}

\begin{figure*}
    \centering
    \includegraphics[width=0.95\linewidth]{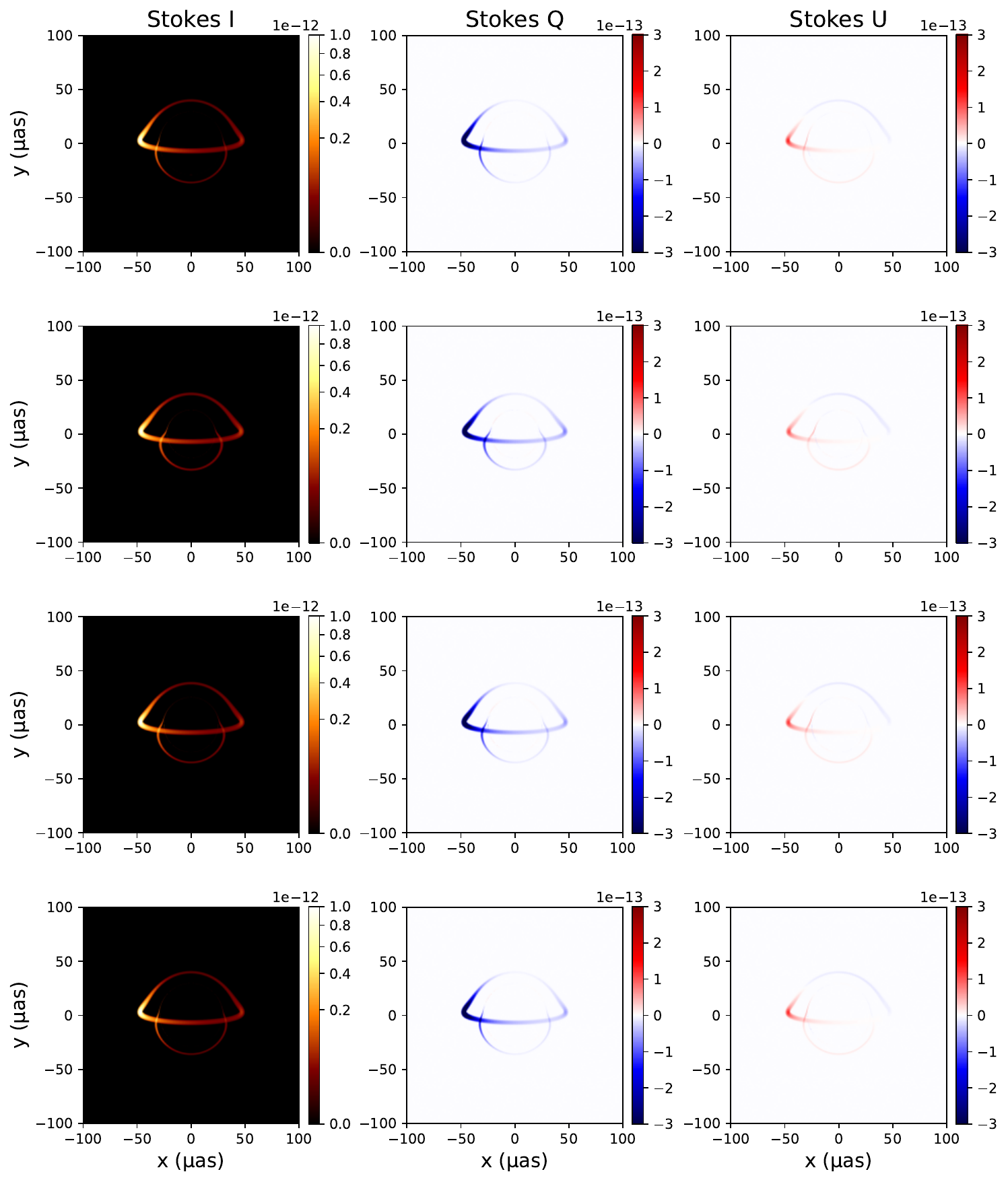}
    \caption{Integrated flux for the Stokes $I$ (left column), Stokes $Q$ (middle column), and Stokes $U$ (right column) parameters, for an observation inclination of $80^\circ$, for the negative JP model (first row), positive JP model (second row), negative KZ model (third row), and positive KZ model (fourth row).}
    \label{fig:integrated_80}
\end{figure*}

The results of the simulations are organized as follows. In Fig. \ref{fig:integrated_SBH} we present the time integrated fluxes for the Schwarzschild BH. In Figs. \ref{fig:integrated_20} and \ref{fig:integrated_80} we present the time integrated fluxes for the parametrized BH solutions introduced in the previous section for observation inclinations of $20^\circ$ and $80^\circ$, respectively. In Fig. \ref{fig:mag_cent} we present the temporal centroid and temporal magnitude for the Stokes parameter $I$, and the total temporal fluxes for the stokes parameters $Q/I$ and $U/I$. In Fig. \ref{fig:QU_EVPA} we present the temporal QU-loops and temporal EVPA. Finally, in Fig. \ref{fig:QU_time}, we present the time evolution of the points in the QU-plane.

Overall, we observe that the observational properties, both astrometric and polarimetric, are qualitatively similar throughout all solutions analyzed. This is to be expected since the spacetime structure of all solutions is qualitatively similar, i.e., all solutions present a single photon sphere of unstable bound geodesics and an event horizon. Consequently, deviations between the observational properties of different solutions are purely quantitative, see Figs. \ref{fig:mag_cent} and \ref{fig:QU_EVPA}, and directly associated with the radial position of the critical curves (the projection, on the observer's plane, of the photon sphere) and event horizons. Furthermore, we observe that the deviations from the Schwarzschild BH observables are more prominent in the JP parametrizations than in the KZ parametrizations, and that the direction of the deviation with a variation of the free parameters has an opposite effect on both parametrizations, which is caused by the sign difference in the first non-trivial term in Eqs. \eqref{eq:JPgtt} and \eqref{eq:KZgtt}. In what follows, we analyze these deviations in more detail for each of the observables.

\subsection{Time integrated fluxes}

The time integrated images for the Schwarzschild solution are given in Fig. \ref{fig:integrated_SBH}, whereas the time integrated fluxes for the parametrized BH spacetimes are given in Figs. \ref{fig:integrated_20} and \ref{fig:integrated_80}, for observation inclinations of $20^\circ$ and $80^\circ$, respectively. 

We observe the same qualitative image structure in all figures for the time integrated fluxes. Every image presents three components, a larger primary image corresponding to the photons that are emitted from the source and reach the observer without crossing the equatorial plane, a smaller and thinner secondary image corresponding to the photons that perform half a turn around the central object, crossing the equatorial plane once before reaching the observer, and the hints of a very thin light ring contribution, mostly unresolved, corresponding to the photons that perform at least one full revolution around the central object, i.e., crossing the equatorial plane at least twice before reaching the observer, though the latter is barely visible in the images due to pixelation.

Furthermore, we observe that the intensity and polarization characteristics of the primary images are virtually the same independently of the central object, which implies that any deviations between the other observables are induced by the differences in higher-order images, i.e., secondary and light-ring contributions. While these higher-order contributions also present the same polarization structure, their size and location in the observer's screen vary between different models, which induces the observed deviations. We note also that, for an observation inclination of $80^\circ$, the contribution of the Stokes parameter Q is much larger than the contribution of the Stokes parameter U, which indicates that Q dominates the linear polarization.

\subsection{Temporal centroid, magnitude, and fluxes}

\begin{figure*}
    \centering
    \includegraphics[width=0.98\linewidth]{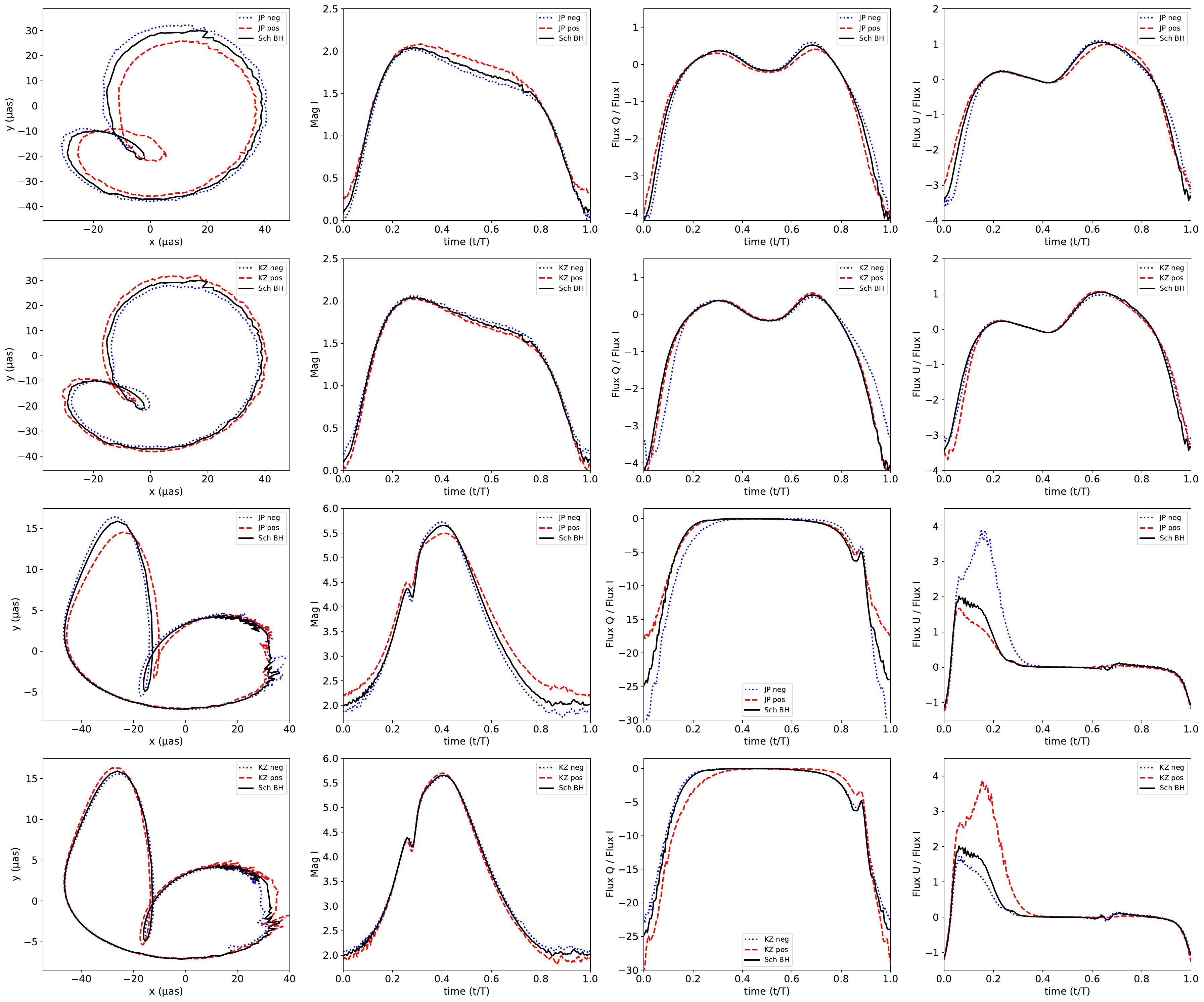}
    \caption{Centroid (left column), magnitude (middle left column), Stokes $Q$ flux (middle right column), and Stokes $U$ flux (right), for the JP models (first and third rows) and KZ models (second and fourth rows), for an observation inclination of $20^\circ$ (top two rows) and $80^\circ$ (bottom two rows).}
    \label{fig:mag_cent}
\end{figure*}

The temporal centroid and temporal magnitude for the Stokes parameter $I$, as well as the total temporal fluxes for the Stokes parameters $Q/I$ and $U/I$, are given in Fig. \ref{fig:mag_cent}. Similarly to what we observe for the integrated fluxes, these observables are qualitatively similar, with deviations among models being purely quantitative. 

We observe that the models that present a secondary image track with a smaller radius in comparison with the Schwarzschild model, namely the positive JP model and the negative KZ model, also present centroid tracks and magnitude values that are smaller than the Schwarzschild case. The inverse is true for the models that present secondary tracks with radii larger than those of the Schwarzschild model. Since the primary images are virtually the same for every model, these results thus indicate that the astrometric observables are dominated by the primary image, as expected due to the fact that the angular size of the primary image is mostly larger than that of the secondary image except when the primary image is dimmed (which happens in the top left part of the orbit), but it is sufficiently affected by the secondary image such that a change in the size of the secondary image radius affects the size of the centroid track. Furthermore, a smaller size of the secondary track implies a larger gravitational bending of light, which is associated with a larger gravitational redshift effect. This explains why the models with smaller secondary image tracks present a lower temporal magnitude, and vice versa.

Our results also indicate that inclination affects particularly the flux of the Stokes parameters $I$ and $U/I$. We observe that the lowest value of the magnitude of $I$ at high inclination is $\sim 2$, whereas at low inclination it was $\sim 0$ by construction, due to the normalization used in the definition of the magnitude. As for the flux of $U/I$, the deviations between the fluxes obtained in Schwarzschild and other models can be large, as happens for the JP negative model and for the KZ positive model, although they also only last for a short interval along the orbit ($t/T\in\left[0.05, 0.40\right]$). For the JP positive and KZ negative models, the largest deviations are also visible at high inclination, but they are less prominent and last for a shorter time period ($t/T\in\left[0.05, 0.25\right]$).

\subsection{Temporal QU-loops and EVPA}

\begin{figure*}
    \centering
    \includegraphics[width=0.98\linewidth]{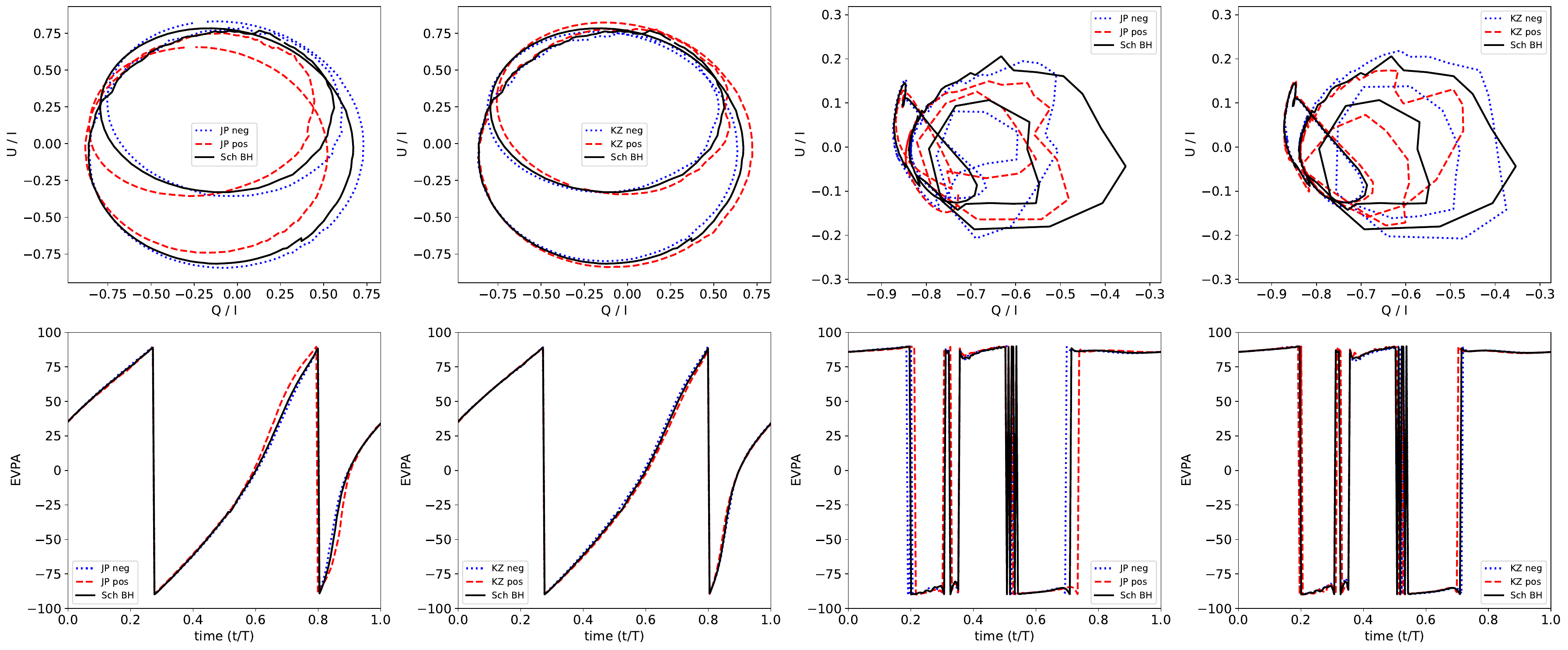}\\
    \caption{QU-loops (top row) and EVPA as a function of time (bottom row) for the JP models (left four panels) and KZ models (right four panels), for an observation inclination of $20^\circ$ (first and second columns) and $80^\circ$ (third and fourth columns).}
    \label{fig:QU_EVPA}
\end{figure*}

\begin{figure*}
    \centering
    \includegraphics[width=0.95\linewidth]{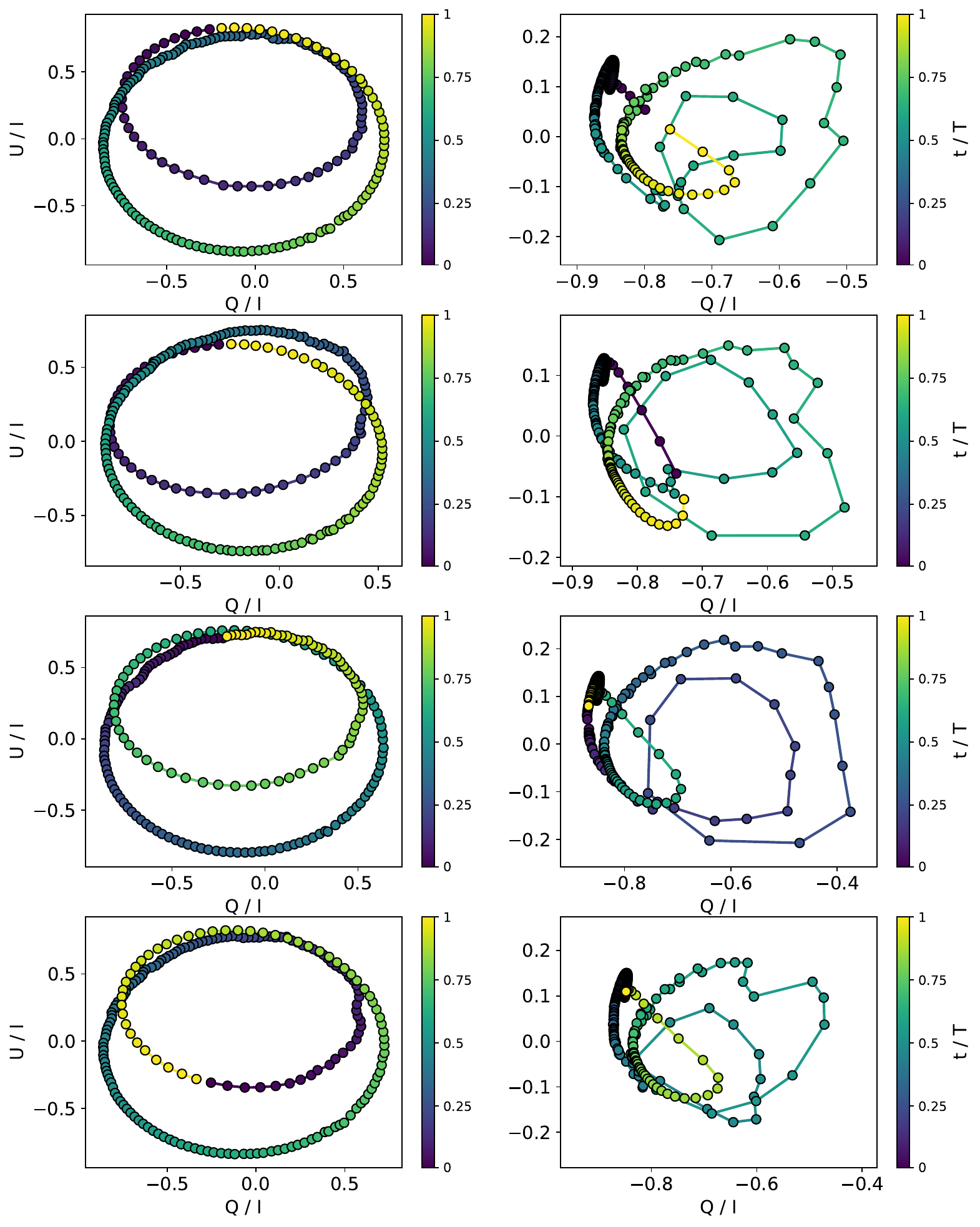}
    \caption{Values in the QU-plane as a function of time for the negative JP model (first row), positive JP model (second row), negative KZ model (third row), positive KZ model (fourth row), for an observation inclination of $20^\circ$ (left column) and $80^\circ$ (right column).}
    \label{fig:QU_time}
\end{figure*}

The temporal QU-loops and temporal EVPA are plotted in Fig. \ref{fig:QU_EVPA}, whereas the time evolution of the points in the QU-plane is given in Fig. \ref{fig:QU_time}. We note that, in the latter figure, the time interval between two consecutive points is constant throughout the figures, which indicates that regions of the plot with a lower density of points correspond to rapidly changing observables, and vice versa. Similarly to what happens in the previous observables analyzed, the deviations among different models are purely quantitative and more prominent for the JP parametrized BHs.

For low inclination, we observe that the tracks in the QU-plane present two loops of different sizes, the smaller one mostly contained within the larger one. This is an expected result whenever a secondary image is present for the vertical magnetic field considered here (for toroidal configurations we would have two loops but with equal radius), and it is consistent with the observations from both ALMA and GRAVITY \cite{GRAVITY:2023avo,Wielgus:2022heh}. The position and orientation of these loops vary slightly among models, with the most prominent deviations from the Schwarzschild model corresponding to the positive JP parametrization model. In what concerns the EVPA, this is also the model that presents the largest deviations with respect to the Schwarzschild solution, but these deviations are again still relatively small and last only for a short time interval in the orbit, $t/T\in\left[0.6, 0.9\right]$, corresponding to the moment the secondary image is beamed (on the left part of the image). As for the KZ parametrization models, the deviations in the EVPA are barely noticeable, which is confirmed from the fact that the loops in the QU-plane are more similar to Schwarzschild in both size and orientation.

For high inclination, the deviations among different models become more prominent. In particular, the right-most part of the tracks in the QU-plane deviates the most. However, as can be seen from the scarcity of points in this region of Fig. \ref{fig:QU_time}, these deviations are short-lived and last only for a small portion of the orbit, whereas the regions with the largest density of points are more qualitatively similar throughout different models. However, there are some noticeable qualitative deviations between the JP parametrization models and the Schwarzschild model at high inclination in the regions where the points in the QU-plane are slowly varying. In particular, one observes that, for the positive JP model, the QU-track presents a crossing on the left-side, which is absent in the negative JP model. Both of these features are qualitatively different in comparison to the ones observed for the Schwarzschild model, which is qualitatively similar to those of the KZ parametrization models. Regarding the EVPA, its complicated structure with multiple jumps caused by the multiple crossings of the track in the QU-plane through the horizontal $U=0$ axis render this observable practically unusable to distinguish between the models at the resolution considered.

\section{Conclusion and discussion}\label{sec:concl}

In this work we have analyzed astrometric and polarimetric observables produced by the orbit of a spherically symmetric light source orbiting a central parametrized BH model and emitting synchrotron radiation. More concretely, we tested four parametrized BH models, two from the JP family of parametrizations, and two from the KZ family, corresponding to the more extreme models for which the size of the shadow is contained within the $2\sigma$-constraints of the EHT telescope. Our results indicate that the observational properties of the parametrized BH models analyzed deviate only slightly from those of the Schwarzschild solution, both from the astrometric and polarimetric sides. 

Since the intensity and polarization characteristics of the primary images are virtually the same for all models tested, these quantitative differences between the models arise from differences in the properties of the secondary and higher-order images, namely their size and position on the observer's screen. Configurations with larger secondary images present wider centroid tracks and higher temporal fluxes, the latter caused by a lower gravitational redshift effect, and vice versa. The astrometric observables extracted from the parametrized JP models present larger deviations with respect to the Schwarzschild model than those extracted from the parametrized KZ models, with the JP model with a positive parameter showing the larger deviations, particularly at large inclinations.

As for the QU-loops one again observes that the parametrized JP models present the largest deviations with respect to the Schwarzschild model. In particular, for the QU-loops, one observes not only deviations in the size of the two loops at low inclination, but also a tilt in the orientation of the loops. At high inclination, one observes a clear qualitative distinction between the QU-tracks obtained for the parametrized JP models and the remaining models, with the left-most part of the tracks (i.e., for larger negative $U$) showing a clear crossing for the positive JP model, which is absent for the negative JP model. Furthermore, the size and shape of the larger loops at high inclination are qualitatively different for all models analyzed, but these differences are only visible for a short fraction of the whole orbit, which may prove difficult to observe experimentally. The temporal EVPA was shown to be the least interesting observable of all, because the deviations between different models are the smallest at low inclination, and because the number of transitions at high inclination is large and makes the interpretation of this observable quite challenging for the resolution considered.

Summarizing, our results indicate that, although the observables extracted from all of the models analyzed present deviations with respect to those extracted from the Schwarzschild spacetime, these deviations may be challenging to observe experimentally, either because they are purely quantitative and small in magnitude, or because the qualitative differences that arise are visible only for a brief fraction of the hot-spot orbit. Nonetheless, this is relevant, since any quantity that is not degenerate over the orbit can be less degenerate with astrophysical parameters. Given the fact that parametrized BHs of this kind are general enough to encompassed large families of metrics, we thus argue that more precise observations are needed in order to experimentally test and constrain BH models this way, and that perhaps the next generation of experiments such as e.g. GRAVITY+ and ngEHT will be up for the challenge. Complementary to this, one could carry out combined measurements of time-averaged images and hot-spots, given the fact that the theoretical quantities characterizing each phenomenon are all related to the critical exponents of the photon sphere, as described in \cite{Kocherlakota:2024hyq}.

\begin{acknowledgments}
This work is supported by the Spanish National
Grants PID2022-138607NBI00 and  CNS2024-154444, funded by MICIU/AEI/10.13039/501100011033 (``ERDF A way of making Europe" and ``PGC
Generaci\'on de Conocimiento") and FEDER, UE.
\end{acknowledgments}


\end{document}